\documentclass[aps,prd,nofootinbib]{revtex4}

\usepackage{graphicx}

\usepackage{epsfig}
\usepackage{color}

\usepackage{amsmath}

\usepackage{amssymb}

\begin{document}

\title{Maximum force conjecture in Kiselev, $4$D-EGB and Barrow corrected-entropy black holes}
\author{K. Atazadeh}\email{atazadeh@azaruniv.ac.ir}

\affiliation{Department of Physics, Azarbaijan Shahid Madani University, Tabriz, 53714-161 Iran}

\date{\today}

\begin{abstract}
The classical maximum force bound in the general relativity (GR) is defined between two black holes with touching horizons. We consider the maximum force conjecture for Kiselev solution that the black holes surrounded by quintessential matter, $w=-2/3$. We show that the maximum force bound is independent of black hole masses in this solution and we also indicate that when two black holes surrounded by static quintessence, the maximum force between them can approach to zero. In continue, we also study the maximum force bound for $4$D Einstein-Gauss-Bonnet ($4$D-EGB) black holes and we obtain that in this theory the maximum force bound exists and the force is bigger than the maximum force in GR. Finally, we consider the Barrow entropy in the framework of the entropic force theories and find that the maximum force only holds when the exponent of the corrected-entropy, namely $\Delta$, goes to zero and for other ranges of $\Delta$ it does not hold in which the mass dependence in the maximum force bound may cause the formation of naked singularities.
\end{abstract}

\maketitle
\date{}
\section{Introduction}
The Maximum force or maximum tension conjecture in GR has been proposed in 2002 by Gibbons \cite{1}. According to this conjecture, the maximum force limit between two black holes with touching horizons is given by $F_{\rm max }=\frac{c^{4}}{4G}\approx3.026\times10^{43}$N, where $c$ is the velocity of light and $G$ is the Newtonian gravitational constant.
Note that in the Newtonian gravity does not exist any bound on forces, when the distance between two point masses can approach arbitrarily
to zero and the gravitational force can go arbitrarily to infinity \cite{2}. In context of ordinary GR because of formation of an event horizon around these point masses the story is different and the maximum force is emerged which it stands as a force between two Schwarzschild black holes of the same mass touching at their horizons. Even if black hole masses are unequal an inequality guarantees the existence of the maximum force bound \cite{3}.
The maximum force is closely related to maximum power conjecture $ P_{\rm max }=cF_{\rm max }=\frac{c^{5}}{4G}$, so-called Dyson luminosity \cite{dyson}.
This conjecture bounds on maximum possible luminosity in gravitational waves, or even for other forms of radiation as an isolated system can radiate \cite{schiller}, \cite{sperhake}.

The maximum force bound and luminosity limits can be originated from the Planck units when the Planck's constant ($h$) is absent and thus they are fully classical and also exist in the presence of a cosmological constant ($\Lambda$) \cite{JBGib}.
In $N$ spatial dimensions, the Planck force in terms of the fundamental constants $G$, $c$ and $h$ can be written as
\begin{equation}\nonumber
F_{\rm pl}=G^{2/(1-N)}c^{(5+N)/(N-1)}h^{(3-N)/(1-N)},
\end{equation}
from the above equation it can be seen that when we choose $N=3$ the Planck's constant, $h$, disappears in $F_{\rm pl}=\frac{c^{4}}{G}$ and it seems that $3$-dimensional space is unique and this may tell something fundamental about these non-quantum natural units in $N=3$. Also, an extension to the magnetic moment to angular momentum has been studied in \cite{BarGib2}.\\
According to Ref. \cite{3}, the maximum force conjecture can be related to formation of naked singularities in $N>3$ dimensions. In $N$ dimensions the quantity $M\times A^{N-2}$ is the classical Planck unit ($h$-free) and by choosing $N =3$ the force is $M\times A$, where $A$ is the Planck unit of acceleration and $M$ is black hole mass. Therefore, the maximum force between two touching $N$-dimensional black holes can be written as \cite{8}
\begin{equation}\nonumber
F_{N}\propto G^{ \frac{-1}{N-2}} M^{\frac{N-3}{N-2}} c^{\frac{2(N-1)}{N-2}}.
\end{equation}
The above force equation indicates that in the higher dimensional GR, when $N > 3$ the force is depended on mass and it can be arbitrarily large as $M$ increases. This interpretation of conjecture may give us some interesting features about the formation of naked singularities in the higher dimensions.

Recently, maximum force conjecture has been studied in some papers: in the context of modified gravity theories it re-examined for Moffat's theory, Brans-Dicke theory and pure Lovelock gravity \cite{8}. In Ref. \cite{9} the authors consider the maximum force conjecture by black hole thermodynamics and entropic force theories. The counterexamples to the maximum force conjecture has been considered in \cite{10} and also the maximum force conjecture is linked to cosmic censorship conjecture in \cite{11}.\\
In these regards, because of interesting aspects of the maximum force conjecture we are interested considering it in the context of other exact black hole solutions.

The paper is sectionalized as follows. In section II, we consider the existence of a maximum force in black hole solutions that they are surrounded by static quintessence and compare the results with the force for Schwarzschild black holes in GR. In section III, we calculate the maximum force bound for black hole solutions in $4$D-EGB that give the same mass independent maximum force as GR. In section IV, the maximum force conjecture for Barrow entropy in context of entropic force is considered. Finally, we give a conclusion, in section V.

\section{Maximum forces and black holes surrounded by quintessence}

Spherically symmetric static solution for Einstein equations in the presence of the quintessence have obtained by Kiselev and thus black hole solutions  surrounded by quintessential matter have been introduced in \cite{12}. In this section we study the maximum force bound for black holes surrounded by quintessence and compare the results with GR maximum force conjecture. \\
Before to proceed our considerations, let us we have  short comments about origin of the word ``quintessence'' in the cosmology community.
In the context of dark energy models, to consider current cosmic acceleration, there are generally two classes of models. The first class is based on ``modified matter models'' in which the matter sector of the Einstein equations contains an exotic matter source with a negative pressure. The second class is based on ``modified gravity models'' in which geometrical sector of the Einstein equations is modified. The quintessence model is one of the representative modified matter models and it is defined by a canonical cosmological scalar field $\phi$ with a potential $V(\phi)$ responsible for the late-time cosmic acceleration \cite{Caldwell}. \\
Therefore, the word ``quintessence'' does not refer to a scalar field with a time-like gradient in the Kiselev space-time but rather it refers to an exotic matter with a negative pressure \cite{13}. However, the Kiselev space-time have some attractive physical and mathematical properties and we try to consider the maximum force conjecture in this scenario.\\
In the following, we consider the force between two static spherically symmetric black holes that they meet each other in the horizons.

\subsection{Black holes surrounded by the quintessence}

Static spherically symmetric solutions of Einstein equations with the quintessential matter surrounding a black hole is given by in the following line element as

\begin{equation}\label{1}
ds^{2}=h(r)dt^{2}-\frac{dr^{2}}{h(r)}-r^{2}d\Omega ^{2},
\end{equation}

where

\begin{equation} \label{2}
h(r)=1-\frac{r_{g}}{r}-\left(\frac{r_{q}}{r}\right)^{3w_q+1},
\end{equation}

here $r_g = 2GM$,\footnote{We take $c=1$.} $M$ is the black hole mass, $r_q$ is the dimensional normalization constants, $G$ is the Newtonian constant and $w_q$ are the quintessential state parameters. For the case of quintessential matter $w_q = -2/3$, the function  $h(r)$ can be written as

\begin{equation}
h(r)=1-\frac{r_{g}}{r}-\frac{r}{r_{q}}.  \label{3}
\end{equation}
This solution denotes a black hole that it is immersed in the quintessential matter and the event horizons are located on

\begin{equation}
r_{\pm }=\frac{1}{2}(r_q\pm\sqrt{r_{q}^{2}-4r_gr_q}),  \label{4}
\end{equation}%
where $r_{-}$ and $r_{+}$ are inner and outer horizons, respectively. However, for $r_q > 4r_g$, always we have $r_g < r_{-} < r_{+} < r_q$.\\
Note that the special case $r_q=4r_g$ yields a non vacuum solution with only one horizon at $r_h =8GM=4r_g$ and also for $r_q\rightarrow\infty$ from equation (\ref{3}), we can see that the Kiselev quintessence solution is reduced to the Schwarzschild solution whit event horizon at $r_h = r_g$.
\\
In the Kiselev space-time with quintessential matter $w_q = -2/3$, the force between two equal mass black holes ($(1/2)Mh'$) contacting at this radius for each can equally be written as

\begin{equation}\label{66}
F_{q}=\frac{GM^{2}}{r^{2}}-\frac{M}{2r_q},
\end{equation}
where $M$ is the black hole mass.
From  equation (\ref{66}) it can be seen that there are two terms in the force relation in which the first term is gravitational force between two black holes and the second term denotes for quintessence force by minus sign that it is in opposition of gravitational force.
By evaluating $F_{q}$ at the horizon, $r_{+}$,  one gets

\begin{equation}\label{6}
F_{q} = \frac{4 G M^2}{\left(r_q +\sqrt{r_q ^2-8 G M r_q }\right)^2}-\frac{M}{2 r_q }.
\end{equation}

For the case $r_q\geq8GM$, always we have

\begin{equation}
|F_{q}|<F_{\rm GR}=\frac{1}{4G}.
\end{equation}
Thus in this case the magnitude of the maximum force is smaller than in GR.\\
The minimal force $F_q= 0$ corresponds to the degenerate case of $r_q = 8GM$. Interestingly, the force between the black holes, that they are immersed in the quintessence, approaches to zero and the gravitational force exactly equals with the quintessence force.

In the next example, for typical case $r_q = 10GM$ one gets
\begin{equation}
F_{q}= -\frac{1}{10G(1+\sqrt{5})}\Rightarrow|F_q|<F_{\rm GR},  \label{7}
\end{equation}
where the minus sign means the quintessence force is dominated.

To clarify our study we take another value for $r_q$ as $r_q = 16GM$, then we get
\begin{equation}
F_{q}= -\frac{1}{16G(1+\sqrt{2})}\Rightarrow|F_q|<F_{GR}. \label{8}
\end{equation}
Therefore, from the above examples it can be seen that the magnitude of the quintessence force is less than the GR maximum force. As a result, we can conclude that $|F_{q}|<F_{GR}$ for $r_q \geq8GM$. To understand general behavior of the force (\ref{6}), we have depicted $F_{q}$ as a function of $r_q$ in Fig. $1$. From the figure we can see that the force is zero at $r_q=8GM$ and also for large values of $r_q$, we have $F_{q}\rightarrow 0$.

\begin{figure*}[ht]
  \centering
  \includegraphics[width=3in]{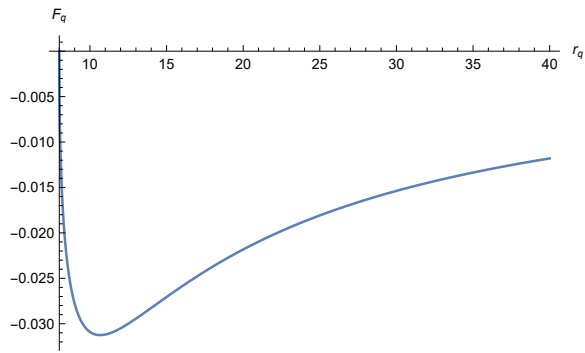}\hspace{1cm}
    \caption{Behavior of $F_{q}$ versus $r_q$. We have set $M=1$ and $G=1$. }
  \label{stable}
\end{figure*}

\subsection{Free quintessence}

By utilizing equation (\ref{2}), the static spherically symmetric free quintessence space-time in the Kiselev solution is given by

\begin{equation}
ds^{2}=\left(1-\frac{r}{r_q}\right)dt^{2}-\frac{dr^{2}}{1-\frac{r}{r_q}}-r^{2}d\Omega ^{2}. \label{9}
\end{equation}

For $-1<w_q<-1/3$, the free quintessence generates the outer horizon of de Sitter kind at $r_{+} = r_q$.

Therefore, the force for free quintessence solution easily can be written as follows

\begin{equation}
F_{fq}=-\frac{M}{2r_q},  \label{10}
\end{equation}
where the force is constant and also from equation (\ref{10}) we see that the gravitational force is absent, because of in the free quintessence space-time there is not any black hole and this constant force only can be generated by quintessential matter $w_q = -2/3$.

For $r_q \geq 2GM$, we have

\begin{equation}
|F_{fq}|\leq F_{GR},  \label{11}
\end{equation}

where the force is smaller than in GR and for $r_q \leq GM$, we get

\begin{equation}
|F_{fq}|> F_{GR}.  \label{12}
\end{equation}
In this case the maximum force conjecture exists in the weak form, {\it i.e.} the coefficient of $1/G$ is bigger than $1/4$. Thus in the free quintessence solution, for the selected values of $r_q$, the force can be bigger or smaller than the GR force.

\section{$4$D-EGB black hole and maximum force}

The Gauss-Bonnet extension to Einstein's gravity in $D\rightarrow 4$ has been proposed in \cite{4D}. However, the validity of this theory is still debatable in view of the fact that the rescaling proposed does not lead to a valid action for $D = 4$ limit of the theory and it seems that internal consistency of the theory is not clear (see for example \cite{4D1} and \cite{4D2}).
Despite such debate in the literatures, this theory evidently admits spherically symmetric static solution and in this case there are two event horizons for vacuum solution. Thus, black holes of this scenario have two horizons.\\
In the following, let us consider the maximum force bound between two $4$D-EGB black holes.
The metric for static spherically symmetric space-time in $4$D-EGB is given by
\begin{equation}
ds^{2}=h(r)dt^{2}-\frac{dr^{2}}{h(r)}-r^{2}d\Omega ^{2},  \label{12}
\end{equation}

where

\begin{equation}
h(r)=1+\frac{r^2}{32 G \pi  \alpha}\left(1\pm\sqrt{1+\frac{128 M \pi  \alpha  G^2}{r^3}}\right) ,  \label{13}
\end{equation}

where $\alpha$ is a finite non-vanishing dimensionless constant in $D=4$. To consider real solution we take $\alpha>0$.

The event horizons are at

\begin{equation}
r_{\pm }=G M\pm\sqrt{G^2 M^2-16 G \pi  \alpha }.  \label{14}
\end{equation}%

So in the $4$D-EGB gravity, the force between two equal-mass black holes at this radius for each one is

\begin{equation}
F_{4D}=\frac{Mr \left(\sqrt{\frac{128 M \pi  \alpha  G^2}{r^3}+1}+1\right)}{32 G \pi  \alpha }-\frac{6 G M^{2}}{2r^2 \sqrt{\frac{128 M \pi  \alpha
   G^2}{r^3}+1}}.
\end{equation}
This equation when evaluated at the horizon, $r_{+}$ can be read as

\begin{align}
F_{4D}=&\frac{\left(G M^{2}+M\sqrt{G \left(G M^2-16 \pi  \alpha \right)}\right) \left(\sqrt{\frac{128 M \pi  \alpha  G^2}{\left(G M+\sqrt{G \left(G M^2-16\pi\alpha \right)}\right)^3}+1}+1\right)}{32 G \pi  \alpha } \\\nonumber&~~~~~-\frac{3G M^{2}}{\left(G M+\sqrt{G \left(G M^2-16 \pi  \alpha \right)}\right)^2 \sqrt{\frac{128 M \pi   \alpha  G^2}{\left(G M+\sqrt{G \left(G M^2-16 \pi  \alpha \right)}\right)^3}+1}}.
\end{align}

For example by taking $\alpha=\frac{GM^{2}}{16\pi}$, we have

\begin{equation}
F_{4D}=\frac{1}{G}>F_{\rm GR}.  \label{15}
\end{equation}
For other values of $\alpha>\frac{GM^{2}}{16\pi}$, we have $F_{4D}>F_{\rm GR}$. So, in this scenario the maximum force conjecture exists in the weak form and the maximum force is bigger than GR force.
To have a general perspective of the force, we have plotted $F_{4D}$ as a function of $\alpha$ in Fig. $2$. Thus, from the plotted figure we can see that the force between black holes for large values of $\alpha$ damps, rapidly.

\begin{figure*}[ht]
  \centering
  \includegraphics[width=3in]{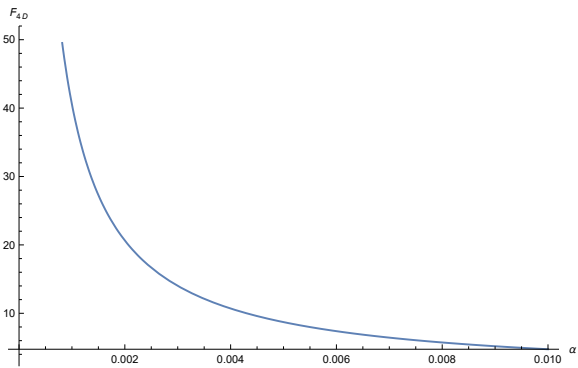}\hspace{1cm}
    \caption{Behavior of $F_{4D}$ versus $\alpha$. We have set $M=1$ and $G=1$. }
  \label{stable}
\end{figure*}

\section{Barrow entropy and maximum force}

Recently, Barrow has proposed that the event horizon of the black holes could be have a pure fractal geometry \cite{15}. The origin of this idea may be linked to quantum-gravitational effects and the fractalized, event horizon can be interpreted in the context of quantum gravity scenarios. Thus, deforming of the event horizon affects the black hole entropy and it can be deviated from the standard Bekenstein-Hawking entropy, as
\begin{equation}
S_B=\left(\frac{A}{A_0}\right)^{1+\frac{\Delta}{2}},
\end{equation}
with $A=4\pi r_h^{2}$ the standard black hole area and $A_0=4G\hbar/k_B$ the Planck area. Hence, the exponent $\Delta$, lying between the extreme values $\Delta = 0$, which corresponds to the standard Bekenstein-Hawking entropy, and $\Delta = 1$, which corresponds to the most fractalized structure of the event horizon.

In this section by taking Barrow entropy, we study existence of maximum force conjecture in the context of the entropic force \cite{16}.
Considerations about the entropic force can be performed for black holes whose Hawking temperature and Bekenstein entropy are given by

\begin{equation}
T=\frac{\hbar \kappa}{2\pi k_B },
\end{equation}
and
\begin{equation}
S_{bh}=\frac{\pi k_B}{G\hbar}r_+^{2},
\end{equation}
where $\kappa$ is the surface gravity of a black hole and $r_+$ denotes for the event horizon of a black hole.

Because $r_+$ is a general event horizon, thus to continue we restrict ourself to charged Reissner-Nordstr\"{o}m black holes in which its surface gravity can be written as \cite{16}
\begin{equation}
\kappa=\frac{1}{r_+^{2}}\sqrt{G^2M^2-\frac{GQ^2}{4\pi\epsilon_0},}
\end{equation}
and the event horizon is
\begin{equation}
r_{+}=GM+\sqrt{G^2M^2-\frac{GQ^2}{4\pi\epsilon_0}},
\end{equation}
where $M$ is the mass, $Q$ is the charge, $\epsilon_0$ is the permittivity of space.
In this way, the Barrow entropy for black holes can be written as
\begin{equation}
S_{B}=\left(\frac{\pi k_B}{G\hbar}\right)^{1+\frac{\Delta}{2}}r_{_+}^{\Delta+2}
\end{equation}

We can obtain the entropic force for $S_{bh}$ as follows

\begin{equation}
F_r=-T\frac{dS_{bh}}{dr_+}=-\frac{1}{G r_+}\sqrt{G^2M^2-\frac{GQ^2}{4\pi\epsilon_0}},
\end{equation}

and for the Barrow entropy we get

\begin{equation}\label{rb}
F_{rB}=-T\frac{dS_{B}}{dr_+}=-(\Delta+2)\left(\frac{\pi k_B }{G\hbar}\right)^{1+\frac{\Delta}{2}}\frac{\hbar}{2\pi k_B}\frac{1}{r_+}\sqrt{G^2M^2-\frac{GQ^2}{4\pi\epsilon_0}}~~r_+^{\Delta}.
\end{equation}
By considering $Q \rightarrow 0$ the above equations reduce to the Schwarzschild black hole and the surface gravity
is given by $\kappa=1/4GM$, the event horizon goes to the Schwarzschild radius $r_+ = r_s = 2GM$.
So, in this case the entropic forces reduce to
\begin{equation}
F_r=-T\frac{dS_{bh}}{dr_s}=-\frac{1}{2G}.
\end{equation}
However, for Bekenstein entropy, the maximum force conjecture holds and the entropic force gravitates toward $1/2G$.
Equation (\ref{rb}) for $Q=0$ can be reduced to
\begin{equation}\label{29}
F_{rB}=-T\frac{dS_{B}}{dr_s}=-(\Delta+2)\left(\frac{\pi k_B}{G\hbar}\right)^{1+\frac{\Delta}{2}}\frac{\hbar}{4\pi k_B}~~r_s^{\Delta}.
\end{equation}
For $\Delta=0$ we have $S_B=S_{bh}$, thus $F_r=F_{rB}$.

But by choosing $\Delta= 1$, the most intricate case,
from  equation (\ref{29}) it can be seen that the entropic force diverges to infinity when the mass of a black hole increasing ($F_{rB}$ is proportional
to the horizon radius $r_s \propto M$). Hence, in the case of Barrow entropy when the event horizon is most fractalized, {\it i.e.} $\Delta=1$, and also for ranges $0<\Delta<1$ the principle of maximum force conjecture does not hold and the formation of naked singularities is unavoidable \cite{3,8}.

\section{Conclusions}

In this paper we have considered the maximum force conjecture for Kiselev black hole solution that the black holes surrounded by quintessential matter, $w=-2/3$. We have demonstrated that when two black holes are surrounded by static quintessence, the maximum force between two black holes approaches to zero and the gravitational force exactly equals with the quintessence force at horizons and also we have seen that the maximum force bound is independent of the black hole masses in this solution.
In continue, we have checked the maximum force bound for $4$D-EGB black hole and we have obtained a mass independent maximum force in three space dimensions. In this theory always the maximum force is bigger than in GR.
We have also considered the maximum force in the framework of the entropic force theories for Barrow corrected-entropy black holes. We have obtained that the maximum force only exists for the case $\Delta=0$ and for other values of entropy exponent the conjecture is not valid.\\
The results obtained in the present work are as follows:\\
$\bullet$ In the Kiselev black hole solutions we have found that there is a mass independent maximum force in the three spatial dimensions and in this case the maximum force is smaller than in GR. We have also obtained that for $r_q=8GM$, where the gravitational force exactly equals with the quintessence force, the force between the black holes is zero. This is an interesting point that can be mentioned in this case. Also, in the free quintessence solution for the values of $r_q\geq 2GM$, the maximum force is smaller than in GR and for $r_q\leq GM$ the maximum force is bigger than in GR.\\
$\bullet$ In the $4$D-EGB theory we have also found a mass independent maximum force in $N=3$ spatial dimensions. We have obtained that in this scenario the maximum force bound exists in the weak form, {\it i.e.} the force is bigger than GR force. For example by fixing parameter of this model ($\alpha$) whit $\alpha=\frac{GM^{2}}{16\pi}$, the explicit form of the force is reduced to $F_{4D}=1/G$. \\
$\bullet$ In the corrected-entropy scenario we have obtained that the maximum force bound only exists for $\Delta=0$ and for other entropy exponent ranges ($0<\Delta\leq1$), it depends on mass and thus the naked singularities can be formed. In the naked singularity case the forces between point particles can become arbitrarily large on approach to the singularity \cite{sh}.

\section*{Acknowledgements}

The author would like to thank an anonymous referee for useful and valuable comments which immensely helped in improving the manuscript.

\end{document}